**Higher Mediterranean diet score is associated with longer time between relapses in Australian females with multiple sclerosis**

Hajar Mazahery[1], Alison Daly[1], Ngoc Minh Pham[1], Madeleine Stephens[1], Eleanor Dunlop[1,3], Anne-Louise Ponsonby[2], Ausimmune/AusLong Investigator Group* Lucinda J Black[1,3]

[1] School of Population Health, Curtin University, Perth, WA, Australia

[2] The Florey Institute of Neuroscience and Mental Health, Parkville, Australia

[3] Institute for Physical Activity and Nutrition (IPAN), School of Exercise and Nutrition Sciences, Deakin University, Melbourne, VIC, Australia

* A list of authors and their affiliations appears at the end of the paper.

**Authorship declaration:** There are no conflicts of interest.

**Correspondence:**

Prof. Lucinda Black

Affiliation and address: Institute for Physical Activity and Nutrition (IPAN), School of Exercise and Nutrition Sciences, Deakin University, 221 Burwood Highway, Burwood, Melbourne, VIC Australia

Phone: +61 3 924 45491

Email: Lucinda.black@deakin.edu.au

**Abstract**

A higher Mediterranean diet score has been associated with lower likelihood of multiple sclerosis. However, evidence regarding its association with disease activity and progression is limited. Using data from the AusLong Study, we tested longitudinal associations (over 10 years' follow-up) between the alternate Mediterranean diet score (aMED) and aMED-Red (including moderate consumption of unprocessed red meat) and time between relapses and disability measured by Expanded Disability Status Scale (EDSS) (n=132; 27 males, 105 females). We used covariate-adjusted survival analysis for time between relapses, and time series mixed-effects negative binomial regression for EDSS. After adjusting for covariates, both higher aMED (aHR=0.94, 95%CI: 0.90, 0.99, p=0.009) and higher aMED-Red (aHR=0.93, 95%CI: 0.89, 0.97, p=0.001) were associated with significantly longer time between relapses in females. Whether specific dietary components of a Mediterranean diet are important in relation to relapses merits further study.

**Introduction**

The pathogenesis of multiple sclerosis (MS) is multifactorial, with both genetic and environmental factors, including dietary components and patterns, playing a role in both susceptibility and progression of disability [1, 2]. The Mediterranean diet (high consumption of extra virgin olive oil, vegetables, fruits, cereals, nuts, and pulses/legumes) has been investigated for its health benefits. In people with MS, evidence from cross-sectional studies suggests that a Mediterranean diet is associated with reduced risk of relapse compared to a mixed diet [3], and a higher Mediterranean diet score is associated with lower risk of disability (assessed using the MS Functional Composite score and Patient Reported Outcomes) [4]. We aimed to test the hypothesis that a Mediterranean diet score (with and without moderate consumption of unprocessed red meat) is associated with longer time between relapses and lower levels of disability over 10 years' follow-up.

**Methods**

**Study population**

We examined data from the AusLong Study, a prospective longitudinal cohort study of people followed since their referral after a first clinical diagnosis of central nervous system demyelination (FCD). Detailed methodology is outlined elsewhere [5]. In brief, between 2003 and 2006, 279 participants aged 18-59 years with an FCD were recruited across four regions of Australia and were subsequently reviewed at 5 and 10 years after recruitment. A total of four males and 24 females were lost to follow up by the 10-year review. For the present study, we included participants with clinically isolated syndrome (CIS) just prior to or at recruitment, who had converted to clinically definite MS by the 10-year review, who had dietary intake data and outcome measures at baseline, 5- and 10-year reviews, and who had plausible total energy intakes (males: 800-4000 kcal; females: 500-3500 kcal) at all time points (n=132; 27 males,

105 females). Ethics approval was obtained from the Human Research Ethics Committees of the nine participating institutions as described previously [5]. All participants provided written informed consent.

**Alternate Mediterranean diet score (aMED)**

Dietary intake in the 12 months prior to the study interview at baseline, and 5- and 10-year reviews was assessed using the 101-item Cancer Council Victoria Dietary Questionnaire for Epidemiological Studies version 2 food frequency questionnaire (FFQ) [1]. The alternate Mediterranean diet score (aMED) was calculated as previously described [6]. We also calculated aMED-Red [1], which considers approximately one daily serving (65 g, range 32.5-97.5 g) of unprocessed red meat as part of the Mediterranean diet [7].

**Outcome measures**

Relapses were determined by a clinician using the 2001 McDonald criteria [8]. Starting from CIS, time (number of days) between relapses was calculated as time from onset of one relapse to the onset of the next relapse up to the 10-year review. The Expanded Disability Status Scale (EDSS) was used by a clinician to assess neurological function on a scale from 0 (no disability) to 10 (death).

**Statistical analysis**

We used univariable analyses to identify covariates associated with outcome measures at $p<0.25$. These covariates were then entered into initial multivariable regression models, stratified by sex, with interactions explored and post-estimation tests conducted. Using backward stepwise removal, covariates at p>0.1 were removed and final models were bootstrapped (500 replicates). To test associations with aMED and aMED-Red, we used covariate-adjusted survival analysis for time between relapses, and time series mixed-effects negative binomial regression for EDSS. All final models were bootstrapped (500 repetitions). Data were analysed using Stata version 18.0 (StataCorp, Texas, USA). Statistical significance was set at p<0.05.

**Results**

Baseline characteristics of included participants are presented in Supplementary Table 1a and 1b. After adjusting for covariates and in females only, for each one-unit increase in the aMED or aMED-Red score there was a 6% and 7% increase in time between relapses, respectively (Table 1). There were no statistically significant associations between aMED or aMED-Red and time between relapses in males, nor between aMED or aMED-Red and EDSS in males or females.

**Discussion and conclusion**

Both higher aMED and aMED-Red were associated with a longer time between relapses in females only. The lack of association in males could be partly explained by the low statistical power; however, we cannot rule out different pathological processes among males and females. The positive findings observed in females are in line with the findings of an international observational study, where the risk of relapse in the preceding year in people with MS was 2.6% lower in association with a Mediterranean diet pattern compared to a mixed diet [3].

We found no evidence of an association between aMED or aMED-Red and EDSS, and to our knowledge no other longitudinal observational study has examined that association. Our findings are supported by a recent randomised controlled trial that reported no effect of a modified Mediterranean diet on disability as measured by EDSS (although a positive effect on fatigue was demonstrated) [9].

Our results suggest that moderate consumption of unprocessed red meat is not harmful in relation to time between relapses nor disability measured using EDSS. It is not clear whether any specific dietary components of the Mediterranean diet are particularly important in relation to disease activity. However, there are several candidates such as protein, polyphenols, polyunsaturated and monounsaturated fatty acids, and reduced inflammatory potential. We have previously shown that a pro-inflammatory diet increases relapse rate and FLAIR lesion volume on MRI in people with MS [10].

The main strengths of our study lie in the prospective design of the AusLong Study (where recruitment was at or soon after CIS) and its low loss to follow-up. However, there are some limitations to our study, such as its generalisability (participants were Australian-based and from only four locations), recall bias associated with self-reported dietary assessments, and potential for residual confounding or other unmeasured lifestyle characteristics. For the present study, we included only two markers of MS disease activity/progression, which is also a limitation.

Our results suggest that a Mediterranean diet, with or without moderate consumption of unprocessed red meat, is associated with longer time between relapses in females. Future studies could investigate other disease activity and progression markers to support our findings.

**Acknowledgements:**

We express our heartfelt thanks to the participants in the Ausimmune and AusLong Studies for their time and energy, without which this work would not have been possible. We sincerely acknowledge the outstanding input of research personnel and research officers working on the studies.

The members of the Ausimmune/AusLong Investigators Group are as follows: Robyn M Lucas (National Centre for Epidemiology and Population Health, Canberra), Keith Dear (University of Adelaide, Australia), Anne-Louise Ponsonby and Terry Dwyer (Murdoch Childrens Research Institute, Melbourne, Australia), Ingrid van der Mei, Leigh Blizzard, Steve Simpson-Yap and Bruce V Taylor (Menzies Institute for Medical Research, University of Tasmania, Hobart, Australia), Simon Broadley (School of Medicine, Griffith University, Gold Coast Campus, Australia), Trevor Kilpatrick (Centre for Neurosciences, Department of Anatomy and Neuroscience, University of Melbourne, Melbourne, Australia), David Williams and Jeanette Lechner-Scott (University of Newcastle, Newcastle, Australia), Cameron Shaw and Caron Chapman (Barwon Health, Geelong, Australia), Alan Coulthard (University of Queensland, Brisbane, Australia), Michael P Pender (The University of Queensland, Brisbane, Australia) and Patricia Valery (QIMR Berghofer Medical Research Institute, Brisbane, Australia).

**Declaration of Conflicting Interests:**

The author(s) declare no potential conflicts of interest with respect to the research, authorship and/or publication of this article.

**Funding:**

Funding for the Ausimmune Study was provided by the National MS Society of the United States of America (RG3364A1/2), the National Health and Medical Research Council

(NHMRC) of Australia (GNT313901) and MS Australia. LJB is supported by MSWA and an MS Australia Postdoctoral Fellowship. HM is supported by MSWA. ED is supported by an MS Australia Postdoctoral Fellowship. MP and AD were supported by the Western Australia Department of Health Future Health Research and Innovation (FHRI) Fund. ALP is supported by an NHMRC Investigator Grant.

**Authorship:**

**Hajar Mazahery:** Writing – Original Draft; **Alison Daly**: Methodology, Formal analysis, Supervision, Writing - Review & Editing; **Ngoc Minh Pham:** Writing – Review and Editing; **Madeleine Stephens**: Writing - Review & Editing; **Eleanor Dunlop**: Supervision, Writing – Review & Editing; **Lucinda J Black**: Conceptualization, Funding acquisition, Methodology, Supervision, Writing - Review & Editing

**Table 1:** Associations between aMED and aMED-Red and time between relapses and EDSS in males and females with multiple sclerosis [1]

|  | Time between relapses | | | |
|---|---|---|---|---|
|  | **Male** | | **Female** | |
|  | **aHR (95% CI)** | **p** | **aHR (95% CI)** | **p** |
| aMED [2] | 1.05 (0.91, 1.21) | 0.542 | 0.94 (0.90, 0.99) | 0.009 |
| aMED-Red [2] | 1.06 (0.92, 1.23) | 0.428 | 0.93 (0.89, 0.97) | 0.001 |
|  | EDSS | | | |
|  | **Male** | | **Female** | |
|  | **aIRR (95% CI)** | **p** | **aIRR (95% CI)** | **p** |
| aMED [2] | 0.92 (0.82, 1.04) | 0.180 | 0.97 (0.93, 1.00) | 0.072 |
| aMED-Red [2] | 0.92 (0.82, 1.03) | 0.131 | 0.97 (0.94, 1.01) | 0.198 |

aHR, adjusted hazard ratio; aIRR, adjusted incident rate ratio; aMED, alternate Mediterranean diet score; aMED-Red (aMED including moderate consumption of unprocessed red meat); CI, confidence interval; EDSS, Expanded Disability Status Scale

[1] Analyses were limited to those with clinically isolated syndrome just prior to or at recruitment with plausible energy intake (27 males, 105 females)

[2] Final adjusted models were bootstrapped. Adjusted for covariates remaining statistically significant at p<0.1. Males: adjusted for times the food frequency questionnaire was collected, history of omega-3 supplement use at baseline, and body mass index; Females: adjusted for times the food frequency questionnaire was collected, age, education, history of smoking, history of omega-3 supplement use at baseline, body mass index, use of immunomodulatory medication since last review.

**Supplementary Table 1a:** Characteristics of included study participants[1]

| Characteristics [2] | Male (n=27) | Females (n=105) |
|---|---|---|
| Study region, % (n) | | |
|   Queensland | 18.5 (5) | 32.4 (34) |
|   New South Wales | 22.2 (6) | 18.1 (19) |
|   Victoria | 33.3 (9) | 18.1 (19) |
|   Tasmania | 25.9 (7) | 31.4 (33) |
| Age (years), median (IQR) | 37.5 (12) | 39.0 (15) |
| Education, % (n) | | |
|   Up to year 10 | 22.2 (6) | 22.8 (23) |
|   Year 11 or 12 | 29.6 (8) | 18.8 (19) |
|   TAFE/Trade/Apprentice | 25.9 (7) | 27.7 (28) |
|   University | 22.2 (6) | 30.7 (31) |
| Ever smoked, % (n) | | |
|   No | 37.0 (10) | 39.8 (41) |
|   Yes | 63.0 (17) | 60.2 (62) |
| History of infectious mononucleosis, % (n) | | |
|   No | 63.0 (17) | 65.0 (67) |
|   Yes | 29.6 (8) | 27.2 (28) |
|   Don't know | 7.4 (2) | 7.8 (8) |
| Taking omega-3 supplement at baseline, % (n) | | |
|   No | 85.2 (23) | 80.0 (84) |
|   Yes | 14.8 (4) | 20.0 (21) |
| Taking vitamin D supplement at baseline, % (n) | | |
|   No | 74.1 (20) | 75.2 (79) |
|   Yes | 25.9 (7) | 24.8 (26) |
| Taking disease modifying therapy [3], % (n) | | |
|   No | 40.7 (11) | 31.4 (33) |
|   Yes | 59.3 (16) | 68.6 (72) |
| Relapses, median (IQR) | 3 (1) | 4 (4) |
| BMI (kg/m$^2$), median (IQR) | 27.6 (6.1) | 26.3 (6.8) |
| Physical activity (METS), median (IQR) | 3279 (4902) | 1920 (3126) |
| Hours sitting daily, median (IQR) | 5 (3) | 5 (5) |

BMI, body mass index; CIS, clinically isolated syndrome; EDSS, Expanded Disability Status Scale; IQR, interquartile range; MS, multiple sclerosis; TAFE, Technical and Further Education; METS, metabolic equivalent of ask derived from the International Physical Activity Questionnaire scoring

[1] We included participants with CIS just prior to or at recruitment, who had converted to clinically definite MS by the 10-year review, who had dietary intake data and outcome measures at baseline, 5- and 10-year reviews, and who had plausible total energy intakes (males: 800-4000 kcal; females: 500-3500 kcal) at all time points

[2] Collected at baseline: age, sex, study region, history of infectious mononucleosis, vitamin D supplement use, omega-3 supplement use. Collected at each review: education, total energy intake from food frequency questionnaire, smoking history, use of immunomodulatory medication, hours per day sitting, physical activity, and BMI

[3] Between the 5- and 10-year reviews

**Supplementary Table 1b:** aMED and aMED-Red scores and EDSS scores for included participants [1]

|  | Male | | | Female | | |
|---|---|---|---|---|---|---|
|  | **Baseline** | **5-year review** | **10-year review** | **Baseline** | **5-year review** | **10-year review** |
| Mediterranean diet scores[2] | | | | | | |
|   aMED, mean (SD) | 3.6 (1.7) | 4.1 (1.6) | 4.6 (1.8) | 4.4 (2.2) | 4.3 (2.3) | 4.2 (2.1) |
|   aMED-Red, mean (SD) | 3.4 (1.8) | 4.1 (1.6) | 4.5 (1.7) | 4.4 (2.1) | 4.2 (2.2) | 4.1 (2.1) |
| EDSS[3], median (range) | 1.0 (0.5-1.5) | 2.0 (1.5-2.5) | 2.5 (1.5-3.8) | 1.0 (0.5-1.5) | 1.5 (1.0-2.5) | 2.0 (1.0-3.0) |

aMED, alternate Mediterranean score; aMED-Red (including moderate consumption of unprocessed red meat); EDSS, Expanded Disability Status Scale; SD, standard deviation

[1] We included participants with clinically isolated syndrome just prior to or at recruitment, who had converted to clinically definite MS by the 10-year review, who had dietary intake data and outcome measures at baseline, 5- and 10-year reviews, and who had plausible total energy intakes (males: 800-4000 kcal; females: 500-3500 kcal) at all time points

[2] Mediterranean scoring ranges from 0 (low adherence to the diet) to 9 (maximum adherence to the diet)

[3] EDSS scoring ranges from 0 (no disability) to 10 (death) in 0.5 increments